\documentclass{PoS}
\def\dsigma{{\rm d} \hat\sigma}

\title{Second order QCD corrections to gluonic jet production at hadron colliders}

\ShortTitle{NNLO dijet production}

\author{James Currie\\
        Department of Physics, University of Z\"urich, Winterthurerstrasse 190, 8057 Z\"urich, Switzerland\\
        E-mail: \email{jcurrie@physik.uzh.ch}}

\author{Aude Gehrmann-De Ridder\\
        Institute for Theoretical Physics, ETH Z\"urich, 8093 Z\"urich, Switzerland\\
and\\         Department of Physics, University of  Z\"urich, Winterthurerstrasse 190,  8057 Z\"urich, Switzerland\\
        E-mail: \email{gehra@phys.ethz.ch}}

\author{Thomas Gehrmann\\
        Department of Physics, University of Z\"urich, Winterthurerstrasse 190,  8057 Z\"urich, Switzerland\\
        E-mail: \email{thomas.gehrmann@uzh.ch}}

\author{\speaker{Nigel Glover}\\
        Institute for Particle Physics Phenomenology, University of Durham, South Road, Durham DH1 3LE, England\\
        E-mail: \email{e.w.n.glover@durham.ac.uk}}

\author{Joao Pires\\
Dipartimento di Fisica, Universita di Genova and INFN Sezione di Genova \&
Via Dodecaneso 33, I-16146 Genova, Italy\\ and\\
Dipartimento di Fisica, Universita di Milano-Bicocca \& INFN, Sezione di Milano-Bicocca,
Piazza della Scienza 3, 20126 Milan, Italy\\
        E-mail: \email{joao.pires@mib.infn.it}}

\author{Steven Wells\\
        Institute for Particle Physics Phenomenology, University of Durham, South Road, Durham DH1 3LE, England\\
        E-mail: \email{steven.wells@durham.ac.uk}}

\abstract{We report on the calculation of the next-to-next-to-leading order (NNLO) QCD corrections to the production of two gluonic jets at hadron colliders. In previous work, we discussed gluonic dijet production in the gluon-gluon channel.  Here, for the first time, we update our numerical results to include the leading colour contribution to the production of two gluonic jets via quark-antiquark scattering.  }

\FullConference{ Loops and Legs in Quantum Field Theory - LL 2014,\\
		 27 April - 2 May 2014 \\
		 Weimar, Germany }

\begin{document}

\section{Introduction}

In hadron colliders, the production of high transverse momentum jets is the footprint of fundamental QCD processes.
Single inclusive jet and dijet observables probe the basic 
parton-parton scattering in $2\to 2$ kinematics. Precise measurements of these observables enables 
the determination of both the parton distribution functions in the proton 
and the strong coupling constant $\alpha_s$ up 
to the highest energy scales that can be attained in collider 
experiments. 

Precision measurements of single jet and dijet cross sections
have been performed at the Tevatron~\cite{cdfjet,d0jet} 
 and at the LHC operating at $\sqrt{s}=7$~TeV~\cite{atlasjet7,cmsjet7} and $\sqrt{s}=8$~TeV~\cite{cmsjet8}.
The jet data are frequently included in  
global fits of parton distributions, where they provide crucial 
information on the gluon content of the proton and have been used to determinate the strong coupling by D0~\cite{d0asjet} and CMS~\cite{cmsaspdf}.

In QCD, the (renormalised and mass factorised) inclusive
cross section for a dijet production in proton-proton collisions has the factorised form,
\begin{equation}
\label{eq:totsig}
{\rm d}\sigma =\sum_{i,j} \int   
\frac{d\xi_1}{\xi_1} \frac{d\xi_2}{\xi_2} f_i(\xi_1,\mu_F^2) f_j(\xi_2,\mu_F^2) \dsigma_{ij}(\alpha_s(\mu_R),\mu_R,\mu_F) \nonumber
\end{equation}
where the probability of finding a parton of type $i$ in the proton, carrying a momentum fraction $\xi$, is described by the parton distribution function $f_i(\xi,\mu_F^2)d\xi$ and the partonic cross section ${\rm d}\hat\sigma_{ij}$  for parton $i$ to scatter off parton $j$, normalised to the hadron-hadron flux\footnote{The partonic cross section normalised to the parton-parton flux is obtained by absorbing the inverse factors of $\xi_1$ and $\xi_2$ into $\dsigma_{ij}$.} is summed over the possible parton types $i$ and $j$. As usual $\mu_R$ and $\mu_F$ are the renormalisation and factorisation scales which are frequently set to be equal for simplicity, $\mu_{R}=\mu_{F}=\mu$.

For suitably high centre of mass scattering energies, the infrared-finite partonic cross section has the perturbative expansion 
\begin{equation}
\label{eq:sigpert}
\dsigma_{ij} = {\rm d}\hat\sigma_{ij}^{LO}
+\left(\frac{\alpha_s(\mu_R)}{2\pi}\right)\dsigma_{ij}^{NLO}
+\left(\frac{\alpha_s(\mu_R)}{2\pi}\right)^2\dsigma_{ij}^{NNLO}
+{\cal O}(\alpha_s^3)
\end{equation}
where the next-to-leading order (NLO) and next-to-next-to-leading order (NNLO) strong corrections are identified. The leading-order dijet cross section is proportional to $\alpha_{s}^{2}$. 

Theoretical predictions for dijet observables are available to 
next-to-leading order (NLO) in 
QCD~\cite{NLOQCD} and the 
electroweak theory~\cite{NLOEW}.
The estimated uncertainty from missing higher order corrections 
on the NLO QCD predictions is substantially larger than the 
experimental errors on single jet and dijet data, and is thus the dominant 
source of error in the determination of $\alpha_s$. A consistent inclusion of 
jet data in global fits of parton distributions is currently only feasible at
NLO.  These theoretical limitations to precision phenomenology, coupled with the
spectacular performance of the LHC and LHC experiments, means that next-to-next-to-leading order (NNLO)
accuracy for dijet production is mandatory.

Jets in hadronic collisions can be produced through a variety of different 
partonic subprocesses. The $gg$ channel dominates at the LHC at low $p_T$ whereas at high $p_T$ the
dominant processes are $qq$ and $qg$ scattering. The $qg$ channel has a contribution between 40-50\%
across the whole $p_T$ range making it the second most dominant channel at the LHC. This is not the
case at the Tevatron where $qg$ scattering is the dominant channel at low and moderate $p_T$ and the
high-$p_T$ jet production is completely dominated by $q\bar{q}$ scattering. The first steps towards the NNLO corrections for this process were made in Refs.~\cite{joao4,joao5} which computed the purely gluonic contribution to the dijet cross section, the  $gg \to gg$ subprocess.  In this contribution, 
we provide the first numerical results for the leading colour contribution to the $q\bar q \to gg$ subprocess.  The NNLO calculation presented here describes gluonic jets production in the sense that only $gg \to$~gluons and $q\bar q \to$~gluons matrix elements are involved.

\begin{figure}[t]
(a)\includegraphics[width=4cm]{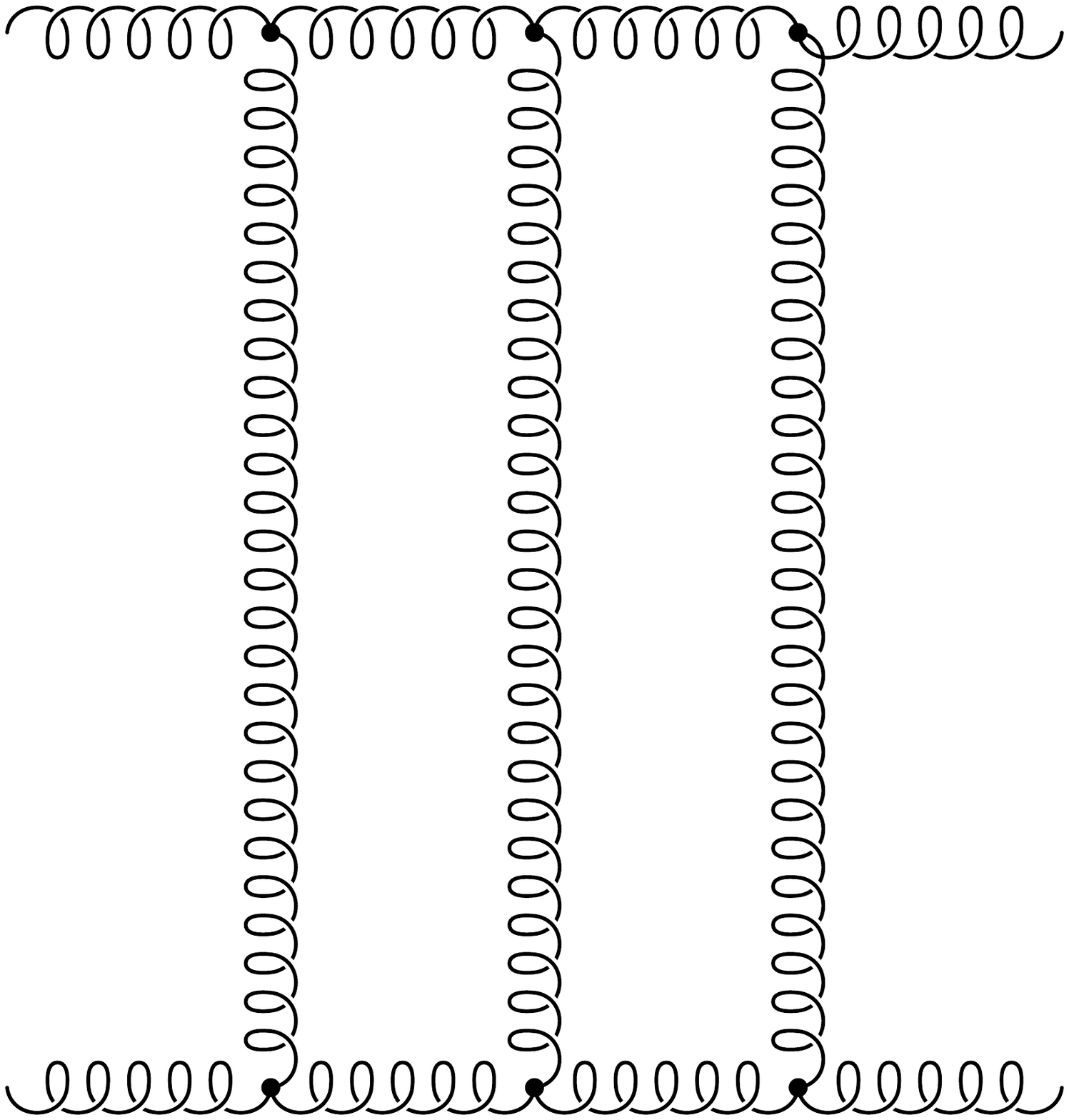}
   \includegraphics[width=4cm]{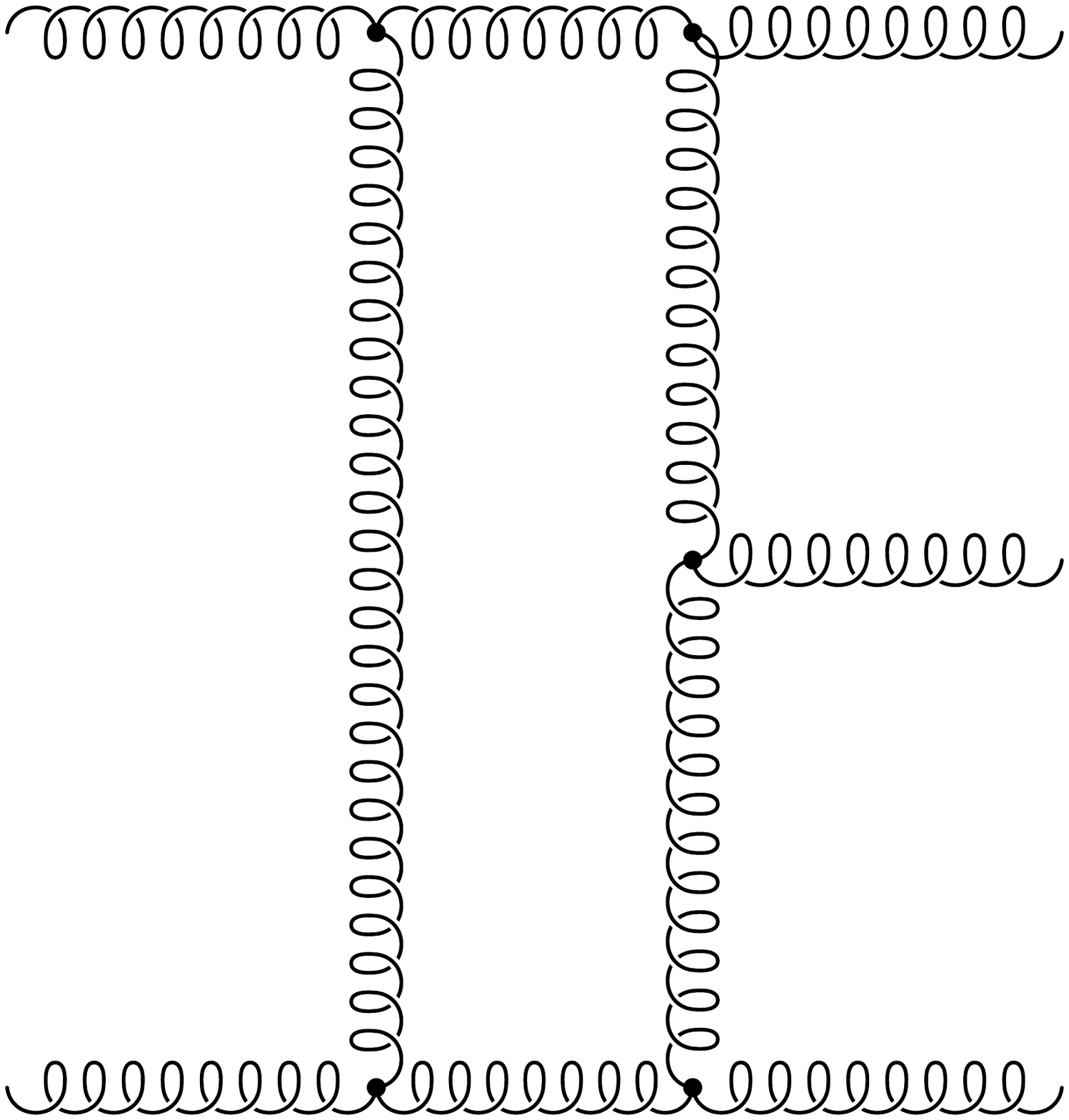}
   \includegraphics[width=4cm]{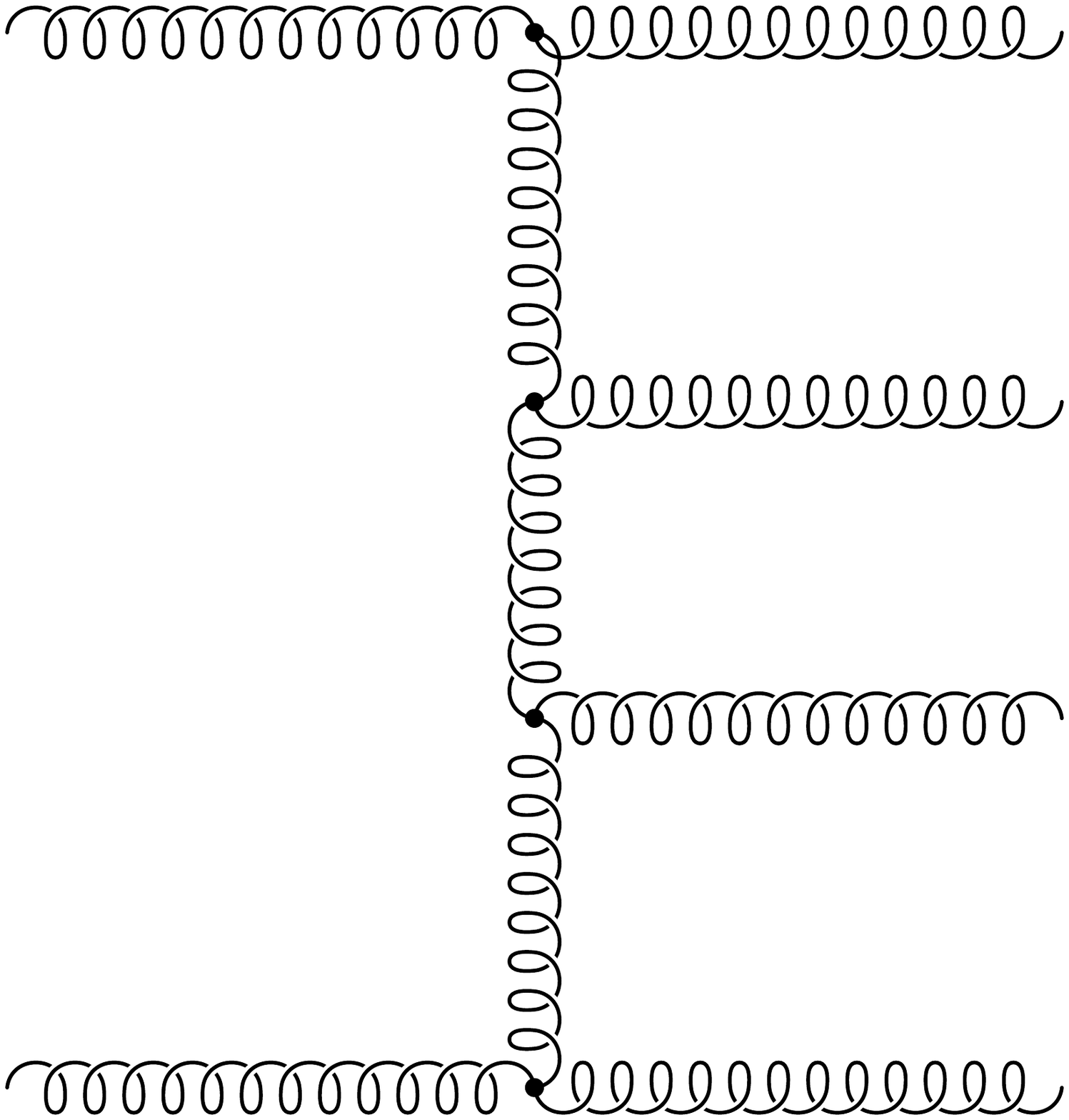}

(b)\includegraphics[width=4cm]{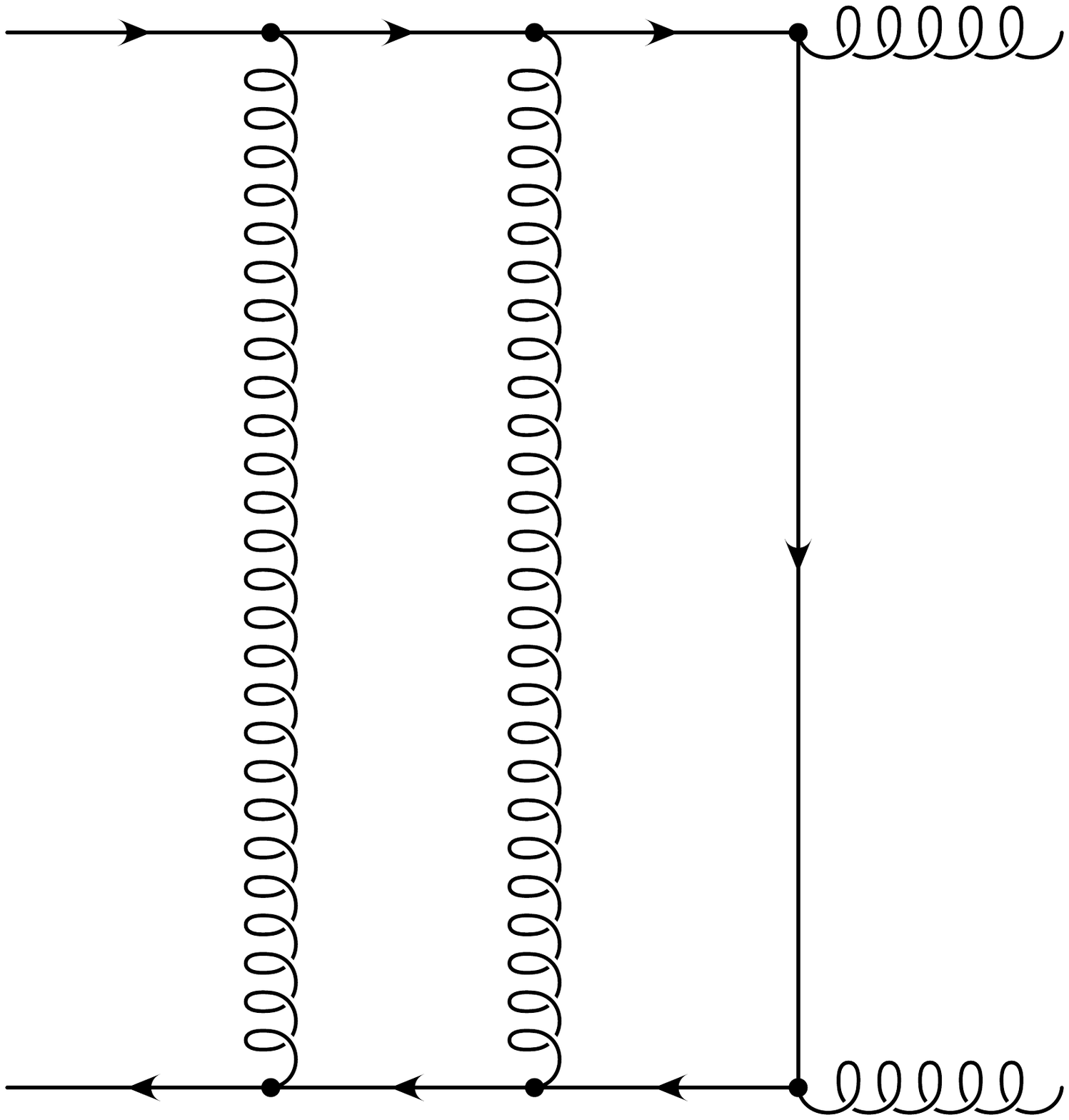}
   \includegraphics[width=4cm]{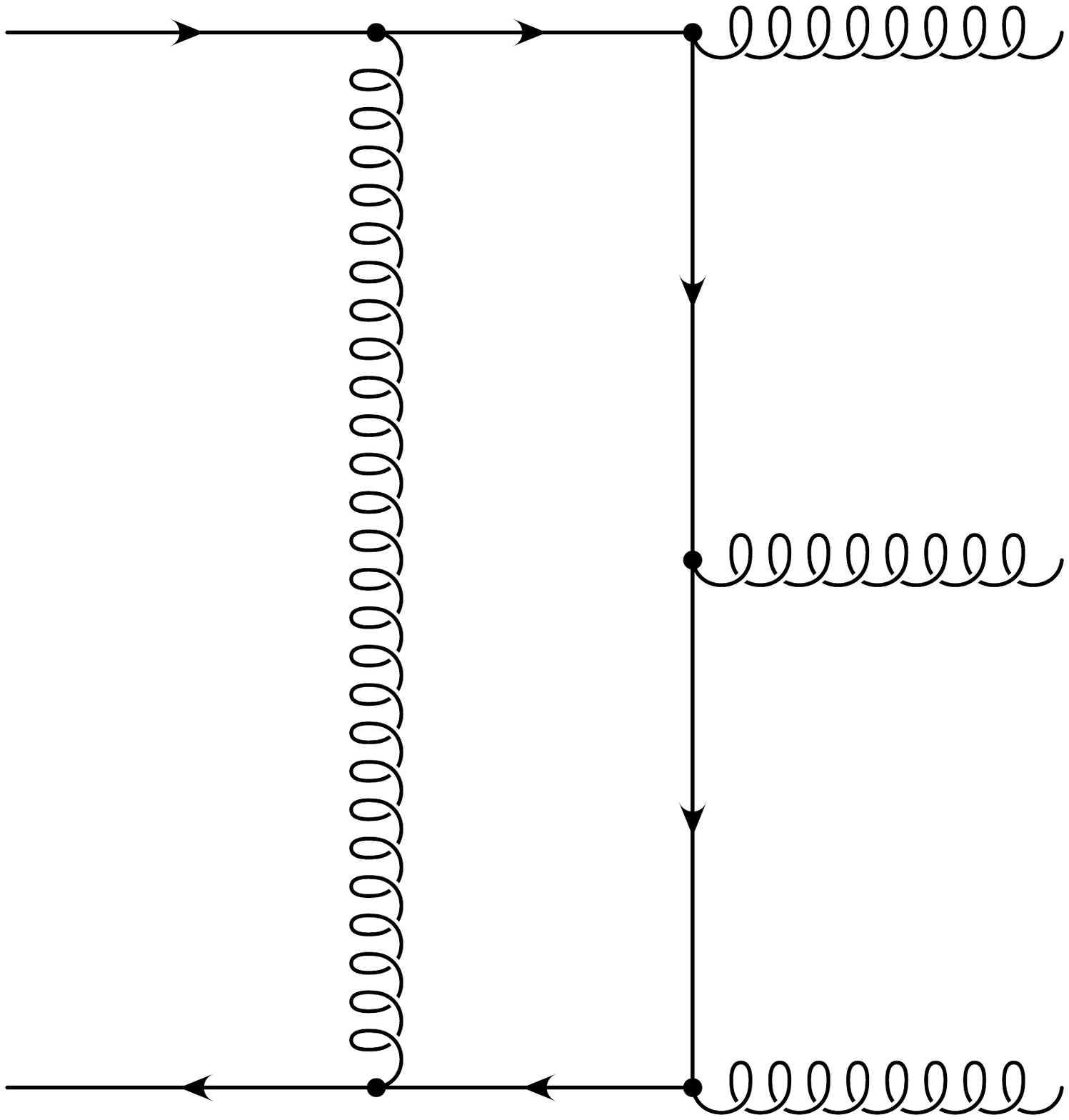}
   \includegraphics[width=4cm]{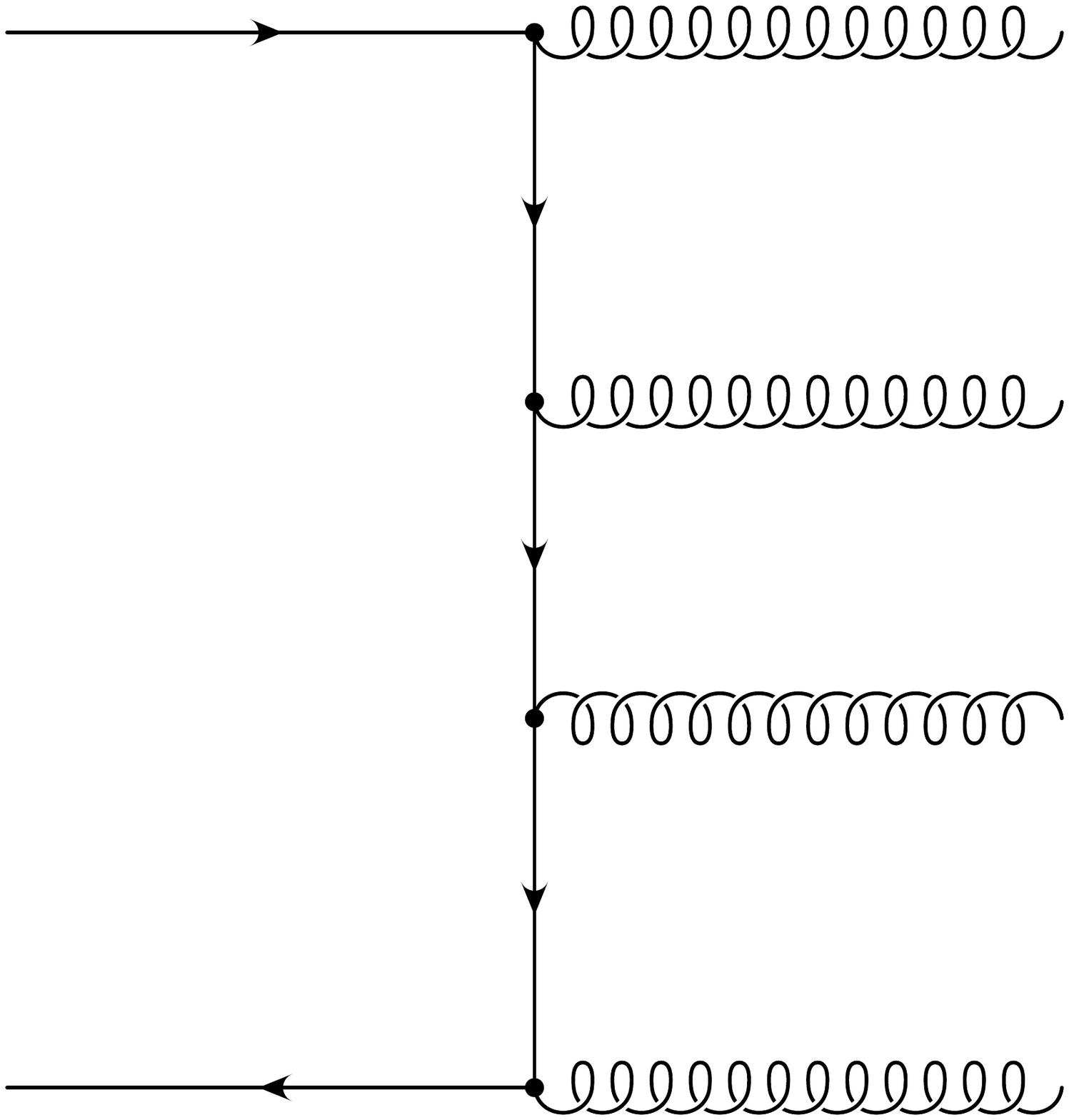}
\caption{Representative Feynman diagrams at NNLO for  
(a) $gg \to$~gluons and (b)
$q\bar{q} \to$~gluons.\label{fig:FD}}
\end{figure}

At NNLO, three types of parton-level 
processes contribute to jet production: the two-loop 
virtual corrections to 
the basic $2\to 2$ process~\cite{twolgggg,twolqqgg}, the one-loop 
virtual corrections 
to the single real radiation $2\to 3$ process~\cite{onelggggg,onelqqggg} and the 
double real radiation $2\to 4$ process at tree-level~\cite{real}. 
Representative Feynman graphs relevant for gluonic dijet production are shown in Fig.~\ref{fig:FD}.

\section{Antenna subtraction and the NNLOJET integrator}

It is well known that in QCD, both the virtual and real radiative corrections are peppered with IR singularities which conspire to mutually cancel to form the finite physical cross section.  
After 
ultraviolet renormalization, the virtual contributions 
contain explicit infrared singularities, which are compensated 
by infrared singularities from single or double real radiation. 
These become explicit only after integrating out the real 
radiation contributions over the phase space relevant to 
single jet or dijet production. This interplay with the jet definition
complicates the extraction of infrared singularities 
from the real radiation process. It is typically done by subtracting 
an infrared approximation from the corresponding matrix elements. These 
infrared subtraction terms are sufficiently simple to be integrated 
analytically, such that they can be combined with the virtual contributions 
to obtain the cancellation of all infrared singularities.
The development of subtraction methods for NNLO calculations is 
a very active field of research and there are several methods on the market: sector decomposition~\cite{secdec}, antenna subtraction~\cite{ourant}, $q_T$-subtraction~\cite{qtsub}
and sector-improved residue subtraction~\cite{stripper} have all been applied successfully in the calculation of NNLO corrections to exclusive processes. 

Here we utilise the antenna subtraction formalism~\cite{ourant,hadant} that was developed for the construction of 
real radiation subtraction terms. It is based on
antenna functions constructed from physical matrix elements~\cite{ourant,ourant2} that each encapsulate all of the infrared singular limits due to unresolved radiation between two hard 
radiator partons.   At NNLO,  antenna functions with up 
to two unresolved partons at tree-level and one unresolved parton 
at one-loop are required. For hadron collider observables, 
one~\cite{gionata} or both~\cite{ritzmann,monni} radiator partons can be in the initial state.  

Our parton-level integrator, NNLOJET, can compute any
infrared-safe  observable related to gluonic dijet final states. NNLOJET  is based around three integration
channels, each identified by the multiplicity of the final state:
\begin{eqnarray}
\dsigma_{ij,NNLO}&=&\int_{{\rm{d}}\Phi_{4}}\left[\dsigma_{ij,NNLO}^{RR}-\dsigma_{ij,NNLO}^S\right]
\nonumber \\
&+& \int_{{\rm{d}}\Phi_{3}}
\left[
\dsigma_{ij,NNLO}^{RV}-\dsigma_{ij,NNLO}^{T}
\right] \nonumber \\
&+&\int_{{\rm{d}}\Phi_{2}}\left[
\dsigma_{ij,NNLO}^{VV}-\dsigma_{ij,NNLO}^{U}\right].
\end{eqnarray}
For each choice of initial state partons $i$ and $j$,  each of the square brackets is finite and well 
behaved in the infrared singular regions. 
For gluonic dijet production there are two channels, $gg \to $~jets and $q\bar q \to $~jets.
The construction of 
the three subtraction terms $\dsigma_{gg,NNLO}^{S,T,U}$ was described at leading colour in 
Refs.~\cite{joao1,joao2,joao3} and at sub-leading colour in Ref.~\cite{joao5} while the leading colour subtraction terms for the $q\bar q \to$~gluons process 
were presented in Ref.~\cite{james1}. It is a feature of the antenna subtraction method that the 
explicit $\epsilon$-poles in the dimensional regularization parameter of one- and two-loop matrix elements are cancelled analytically against the $\epsilon$-poles of the integrated antenna subtraction terms, while the implicit infrared poles present in the singular regions of the double-real and real-virtual phase space cancel numerically.

\section{Numerical results}

As in Refs.~\cite{joao4,joao5}, our numerical studies are based on proton-proton collisions at centre-of-mass energy $\sqrt{s}=8$~TeV.  We focus on the
single jet inclusive cross section 
(where every identified jet in an event that passes the selection cuts 
contributes, such that 
a single event potentially enters the distributions multiple times) 
and the two-jet exclusive cross section (where events
with exactly two identified jets contribute). 

Jets are identified using the anti-$k_T$ algorithm with resolution 
parameter $R=0.7$. Jets are accepted at central rapidity $|y|<4.4$, and 
ordered in transverse momentum. An event is retained if the leading 
jet has $p_{T1}>80$~GeV.  For the dijet invariant mass distribution, a second jet must be observed with $p_{T2}>60$~GeV.

\begin{figure}[th]
  \centering
    \includegraphics[width=0.8\textwidth]{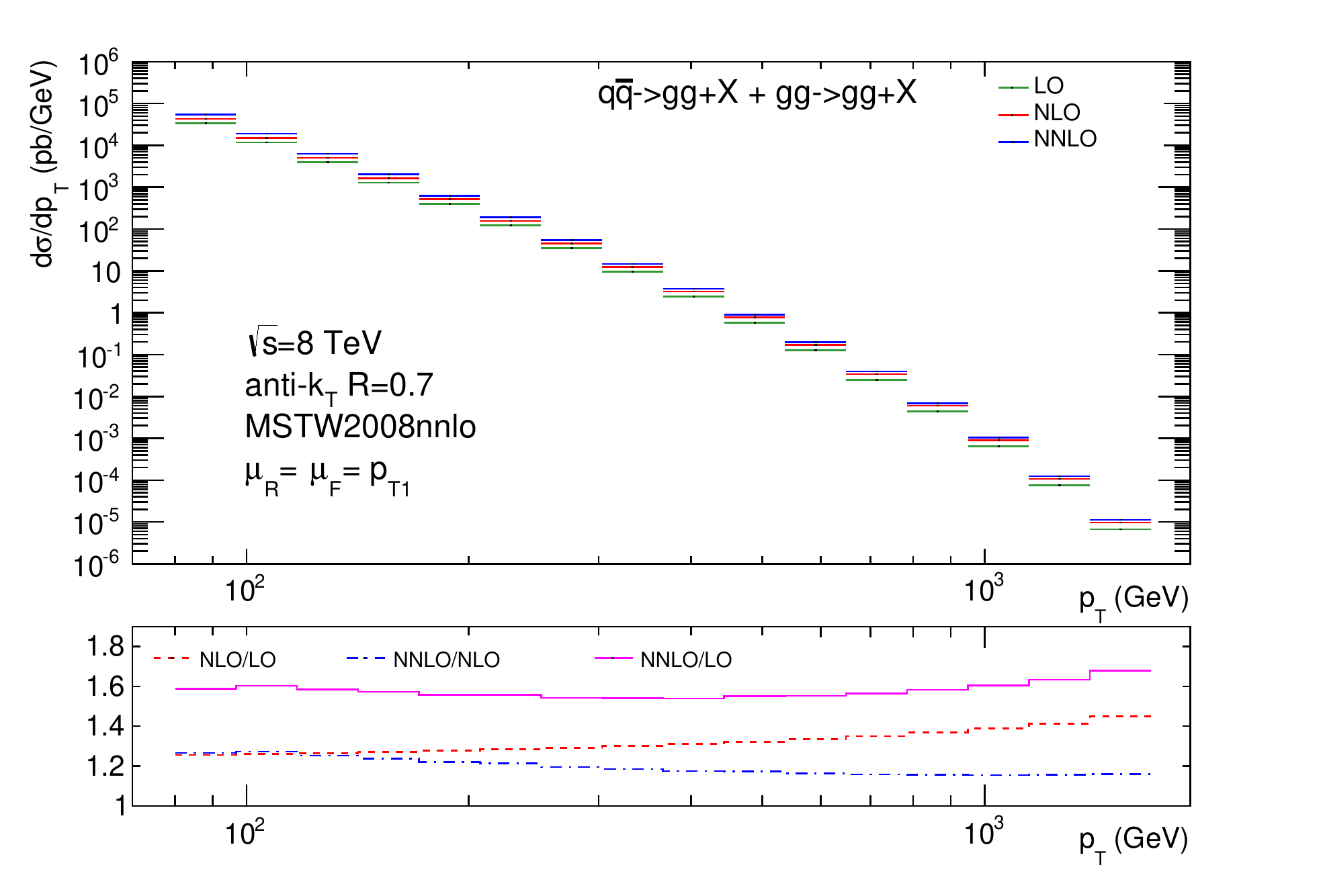}
  \caption{Inclusive jet transverse energy distribution, $d\sigma/dp_T$, for jets constructed with the anti-$k_T$ algorithm with $R=0.7$ and with $p_T > 80$~GeV, $|y| < 4.4$ and $\sqrt{s} = 8$~TeV at NNLO (blue), NLO (red) and LO (dark-green). The lower panel shows the
ratios of different perturbative orders, NLO/LO, NNLO/LO and NNLO/NLO.}
  \label{fig:dsdet}
\end{figure}

All calculations have been carried out with the MSTW08NNLO 
distribution functions~\cite{mstw}, including the evaluation of the 
LO and NLO contributions. 
This choice of parameters
allows us to quantify the size of the genuine NNLO contributions to the  
parton-level subprocess. Factorization and renormalization scales
($\mu_F$ and $\mu_R$)   
are chosen dynamically on an event-by-event basis. As default value, we
set $\mu_F = \mu_R \equiv \mu $ and set $\mu$ equal to the transverse momentum of the leading jet so that $\mu = p_{T1}$. 

In Fig.~\ref{fig:dsdet} we present the inclusive jet cross section for the
anti-$k_T$ algorithm with $R=0.7$ and with $p_T > 80$~GeV, $|y| < 4.4$ as a
function of the jet $p_{T}$ at LO, NLO and NNLO, for the central scale choice
$\mu = p_{T1}$.  The NNLO/NLO $k$-factor shows the size of the higher order NNLO
effect to  the cross section in each  bin with respect to the NLO calculation.
For this scale choice we see that the NNLO/NLO $k$-factor is approximately flat
across the $p_{T}$ range corresponding to a 27-16\% increase compared to the NLO
cross section.  Note that in the combination of $q\bar q \to gg$ +$gg \to gg$ channels, the gluon-gluon initiated channel dominates. The NNLO/NLO $k$-factor for the $q\bar q \to gg$ channel alone is roughly 5\%.   


\begin{figure}[th]
  \centering
    (a)\includegraphics[width=0.46\textwidth]{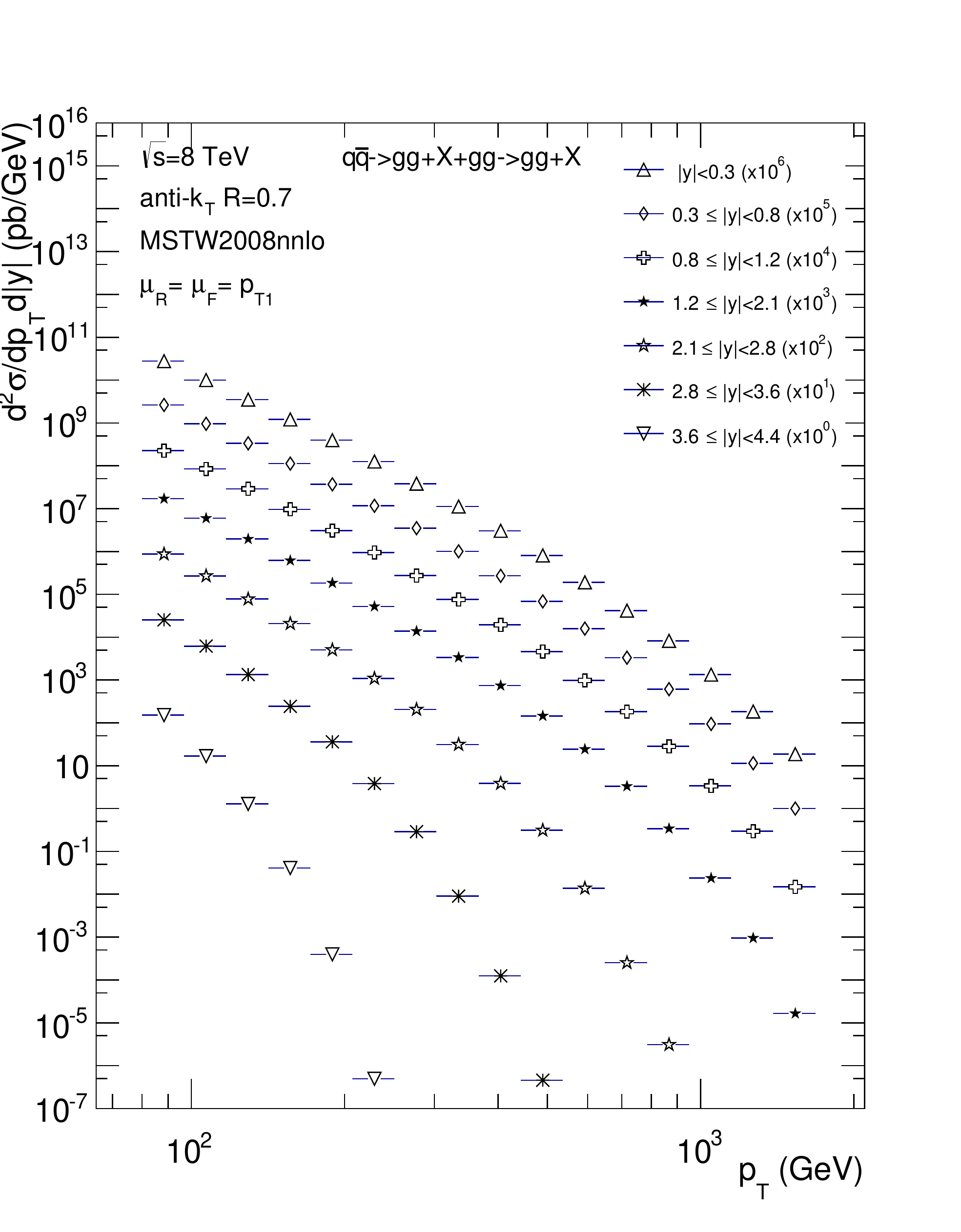}
    (b)\includegraphics[width=0.46\textwidth]{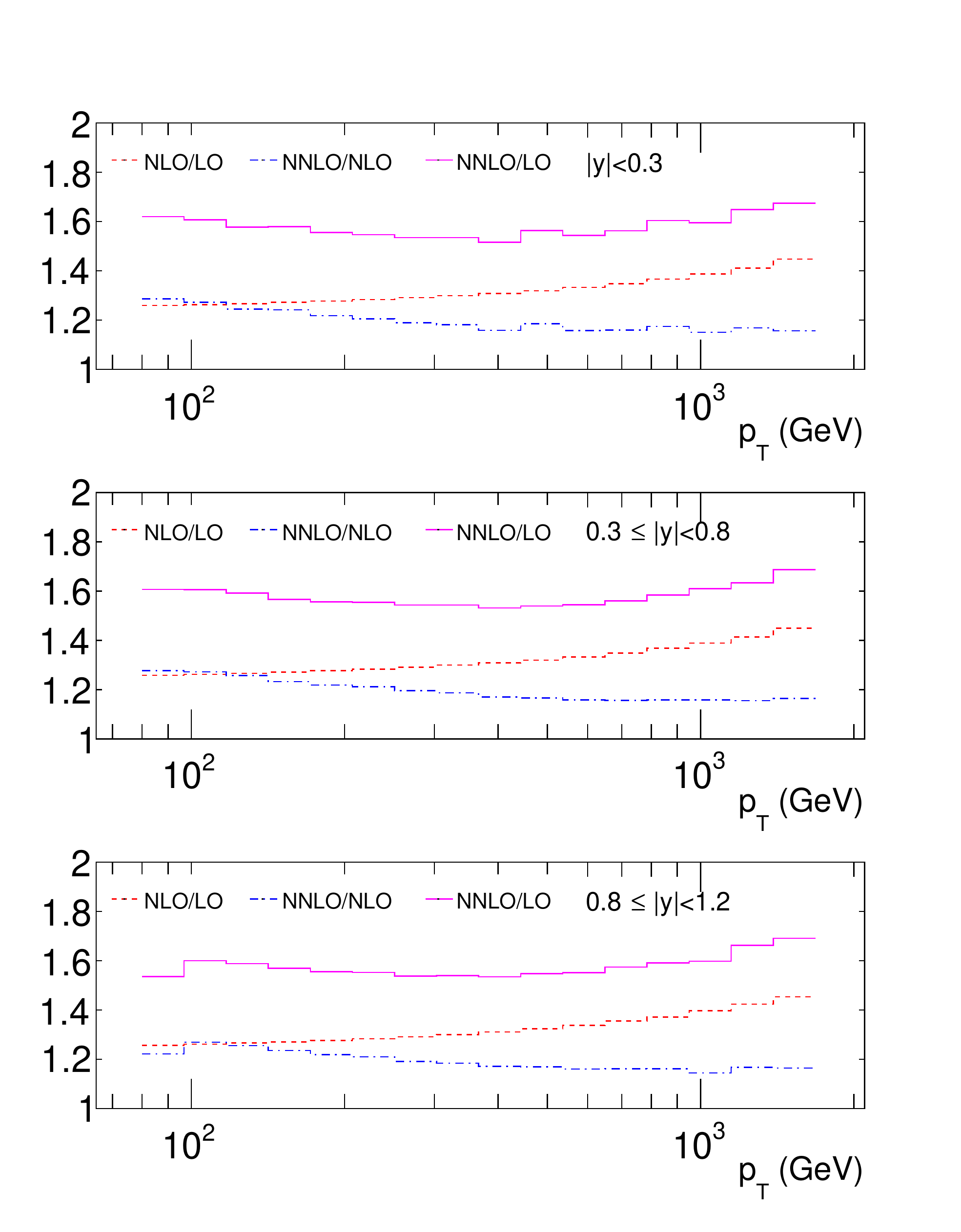}
  \caption{(a) The doubly differential inclusive jet transverse energy distribution, $d^2\sigma/dp_T d|y|$, at $\sqrt{s} = 8$~TeV for the anti-$k_T$ algorithm with $R=0.7$ and for $E_T > 80$~GeV and various $|y|$ slices and (b) double differential $k$-factors for $p_T > 80$~GeV and three $|y|$ slices: $|y | < 0.3$, $0.3 < |y| < 0.8$ and $0.8 < |y| < 1.2$.}
  \label{fig:d2sdetslice}
\end{figure}

Fig.~\ref{fig:d2sdetslice}(a) shows the inclusive jet cross section in double-differential form in
jet $p_{T}$ and rapidity bins at NNLO.  The $p_{T}$ range is divided into 16 jet-$p_{T}$ bins and seven rapidity intervals over the range 0.0-4.4 covering central and forward jets. The double-differential $k$-factors for the distribution in Fig.~\ref{fig:d2sdetslice}(a) for three rapidity slices: $|y | < 0.3$, $0.3 < |y| < 0.8$ and $0.8 < |y| < 1.2$ are shown in Fig.~\ref{fig:d2sdetslice}(b).
We observe that the NNLO correction increases the cross section between 27\% at low $p_{T}$ to 16\% at high $p_{T}$ with respect to the NLO calculation (blue dot-dashed line) and this behaviour is similar for all three rapidity slices.

\begin{figure}[th]
  \centering
   (a)\includegraphics[width=0.46\textwidth]{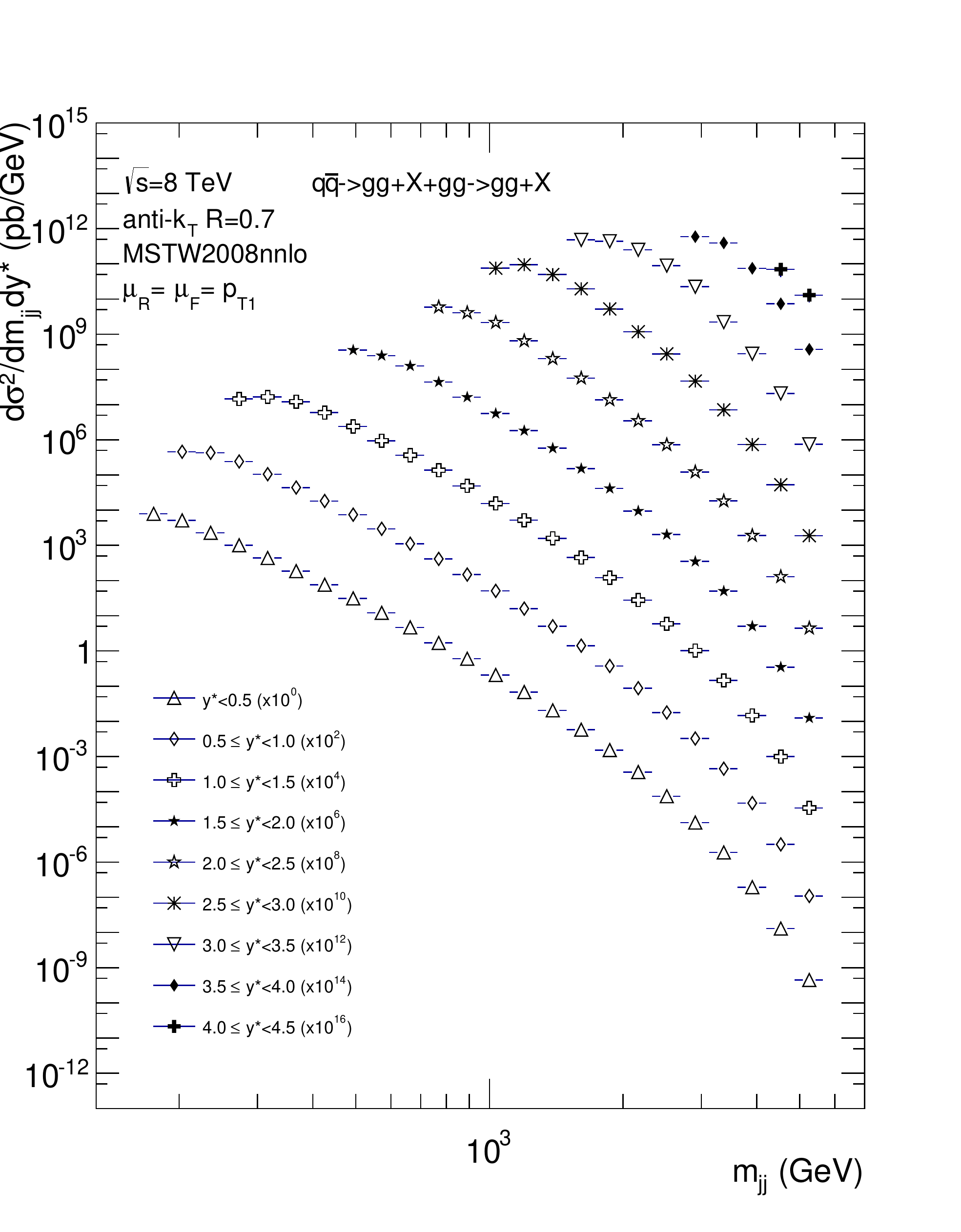}
   (b)\includegraphics[width=0.46\textwidth]{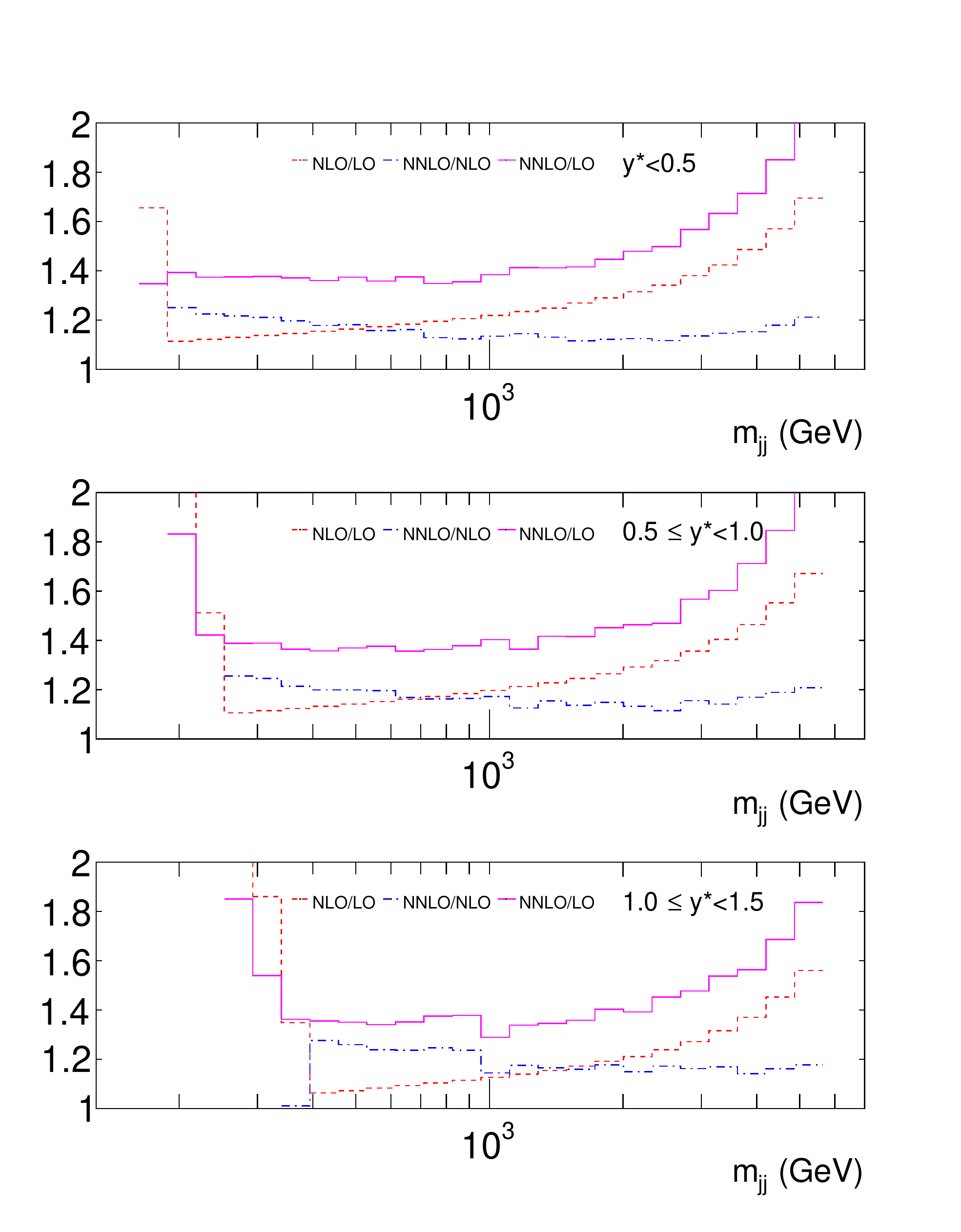}
  \caption{(a) Exclusive dijet invariant mass distribution, $d\sigma/dm_{jj}dy^*$, at $\sqrt{s} = 8$~TeV for $y^* < 0.5$ with $p_{T1} > 80$~GeV, $p_{T2} > 60$~GeV and $|y_1|,~|y_2| < 4.4$ at NNLO (blue), NLO (red) and LO (dark-green) and (b) the
ratios of different perturbative orders, NLO/LO, NNLO/LO and NNLO/NLO.}
  \label{fig:dsdmjj}
\end{figure}

As a final observable, we computed the dijet cross section as a function of the
dijet mass at NNLO. This is shown in Fig.~\ref{fig:dsdmjj} for the scale choice
$\mu = p_{T1}$ together with the LO and NLO results. The dijet mass is computed
from the two jets with the highest $p_{T}$ and $|y_1|,~|y_2| < 4.4$ with $y^{*}$, defined as half the
rapidity difference of the two leading jets $y^{*}=|y_{1}-y_{2}|/2<0.5$.  From Fig.~\ref{fig:dsdmjj}(b), we see
that the NNLO/NLO $k$-factor (blue dot-dashed line) increases the cross section between 25\% at low
$m_{jj}$, 13\% at moderate $m_{jj}$, to 20\% at high $m_{jj}$. Once again this behaviour is similar for all three rapidity slices.

\section{Conclusions}

In conclusion, we have presented numerical results for the fully differential
inclusive jet and dijet cross sections at hadron colliders at NNLO in the
strong coupling constant using the parton-level generator NNLOJET. We have
considered the NNLO QCD corrections to the (full colour) $gg \to$~gluons and 
(leading colour) $q\bar{q}\to$~gluons subprocesses. The remaining contributions including the 
important $qg$ channel are in progress and will be reported on later. 
 

\newpage
\section*{Acknowledgements}
This research was supported in part by
the Swiss National Science Foundation (SNF) under contract 200020-149517, in part by
the UK Science and Technology Facilities Council as well as by the Research Executive Agency (REA) of the European Union under the Grant Agreements PITN-GA-2010-264564 (``LHCPhenoNet''), PITN-GA-2012-316704  (``HiggsTools''), and the ERC Advanced Grant MC@NNLO (340983). JP acknowledges support 
by an Italian PRIN 2010 grant and thanks the Dipartimento di Fisica, Universita di Milano-Bicocca for their kind hospitality.

\end{document}